# Laser Fault Injection in Memristor-Based Accelerators for AI/ML and Neuromorphic Computing

Muhammad Faheemur Rahman[1], Wayne Burleson [2]
[1, 2] Department of Electrical and Computer Engineering, University of Massachusetts Amherst
[1]faheemur@umass.edu; [2]burleson@umass.edu

*Abstract*— Memristive crossbar arrays (MCA) are emerging as efficient building blocks for in-memory computing and neuromorphic hardware due to their high density and parallel analog matrix-vector multiplication capabilities. However, the physical properties of their nonvolatile memory elements introduce new attack surfaces, particularly under fault injection scenarios. This work explores Laser Fault Injection as a means of inducing analog perturbations in MCA-based architectures. We present a detailed threat model in which adversaries target memristive cells to subtly alter their physical properties or outputs using laser beams. Through HSPICE simulations of a large MCA on 45 nm CMOS tech. node, we show how laser-induced photocurrent manifests in output current distributions, enabling differential fault analysis to infer internal weights with up to 99.7% accuracy, replicate the model, and compromise computational integrity through targeted weight alterations by approximately 143%.

## I. INTRODUCTION

The limitations of von Neumann architectures have led to the development of in-memory computing solutions. Memristive crossbar arrays (MCAs) have emerged as a key alternative for energy-efficient analog computation directly within the memory structure. Each memristor stores a weight as a conductance value, and individual column current is determined by summing the product of row voltages and conductances per Ohm's and Kirchhoff's laws. This architecture is inherently parallel and energy-efficient, making it ideal for AI/ML accelerators, neuromorphic computing, and analog edge processors.

Memristors offer non-volatility, retaining their programmed resistance even when powered off, which helps preserve trained neural network weights. However, these advantages come with security risks. The analog nature of MCAs opens up vulnerabilities, especially if an adversary physically accesses the device. By targeting specific cells or control lines with Laser Fault Injection, an attacker can alter or infer the weights stored in the array. Our work develops an attack model demonstrating how LFI-induced perturbations in resistance can expose or corrupt neural network weights. We aim to raise awareness of such threats so future analog accelerators can be designed with stronger fault resilience and model protection.

## II. DEVICE PHYSICS AND LASER FAULT MECHANISM

Memristors are two-terminal devices whose resistance changes in response to the history of voltage or current applied across them. This change occurs due to internal physical effects such as ion migration or the formation of conductive filaments, which modify the electrical path within the device. This property enables memristors to store information as distinct conductance states, representing the strength of neural connections [1]-[2]. Because computational outputs are directly influenced by these resistance values, any unintended shift can degrade accuracy, leaving the system vulnerable to targeted attacks [3].

LFI, traditionally used to trigger bit-flips or timing faults in digital circuits, presents new risks for analog computing platforms like MCAs. In this context, LFI exploits the photoelectric effect within semiconductor substrates to disturb analog storage. When a focused laser beam in the micron range is directed at the silicon region near a memristor or its access transistor, photon absorption generates electron-hole pairs, which in turn produce transient photocurrents. Even brief laser-induced spikes can change the internal state of a memristor, modifying its resistance. In analog domains, small changes can cause noticeable output shifts, posing serious threats to data integrity and system security [5].

## III. ATTACK MODEL AND THREAT SURFACE IN MCA

MCAs consist of orthogonal rows and columns, with memristive devices at each intersection. Each memristor stores a conductance value representing a weight in a neural network. During inference, row voltages are applied, and resulting column currents are read out to represent the product of inputs and weights. The attacker uses a laser beam with a controllable beam size to illuminate and perturb multiple adjacent cells simultaneously, and also moves the laser beam across the array step by step, covering overlapping areas. These actions cause changes in column output currents when laser-induced fault currents vary in magnitude and location, which can then be used for differential fault analysis to infer internal conductances and ultimately extract or corrupt the neural network model.

The threat model includes both passive attacks, which aim to extract stored weights, and active attacks, which intentionally alter conductances to degrade model performance. The overall attack flow is shown in Fig.1. Crucially, we assume the adversary can monitor output currents from the columns but has no direct access to or control over the row inputs, which may be protected using architectural control techniques such as the Keyed Permutor mechanism proposed in [3]. The attacker's only means of influence is through physical fault injection into the crossbar array using laser-based techniques.

## IV. SIMULATION SETUP AND FAULT MODELING

To evaluate the feasibility of Laser Fault Injection (LFI) attacks on memristive crossbar arrays (MCAs), we built a simulation environment using HSPICE. Our test architecture is a 256×128 1T1R crossbar built on 45 nm CMOS process node, where each cell contains a memristor in series with an NMOS selector transistor. To reflect real-world behavior, we included parasitic effects from metal interconnects in the layout. In large

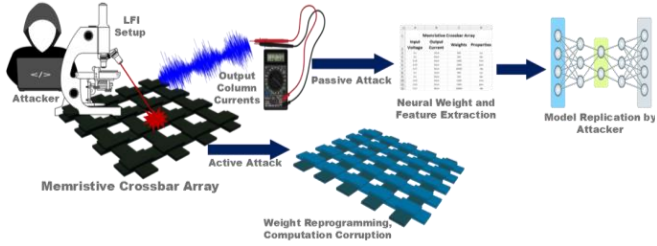

Fig. 1 LFI Adversarial Attack Flow on Memristive Crossbar Array

MCA, it is impractical for an adversary to pinpoint and manipulate individual memristors due to their nanoscale dimensions. Instead, we consider a more realistic adversarial model using a laser with controllable spot or beam size (1–50 µm). In the first experiment, LFI was modeled as a localized current injected at specific points inside the crossbar. We observed how the output current at the columns changed when current faults were introduced. Then, we compared column currents before and after fault injection, modeling memristors as simple linear resistors, with resistance values ranging from 5–20 kΩ. Using these output changes, we applied a linear regression model to estimate the conductance of the fault-affected cells, which corresponds to the internal neural network weights. For the second part of the simulation, we shifted focus to whether injected current could actively alter the stored weights, not just help infer them. For this, we adopted a more realistic memristor model known as the TEAM model [4]. Unlike linear resistance models, TEAM (Threshold Adaptive Memristor) captures the nonlinear and threshold-dependent nature of real memristors. We chose a current-controlled version of the TEAM model because LFI injects current, not voltage. In current-controlled models, the internal state variable of the memristor changes directly as a function of the applied current. This aligns well with laser-induced photocurrent injection. On the other hand, voltage-controlled models would require additional conversions, making them less suitable for LFI-based analysis. We ran experiments to identify whether injected currents from laser fault injection could reliably alter the conductance values of memristors. We also explored current levels needed to induce change, defining attack parameters.

## V. Results and Conclusion

In the first setup, using the HSPICE define crossbar array, we injected various fault currents (10–40 µA) into individual cells across five different resistance values (5 kΩ to 20 kΩ), and the change in output column current was recorded. The results are summarized in Table I, which shows the differential current (ΔI) caused by the injected faults. Taking results from Table I., we draw Fig. 3b which highlights that change in column currents at output increase linearly with the injected current. Using the fault injection data, we applied linear regression to model how the injected current correlates with the change in column current (ΔI) for each memristor resistance. The slope of this relationship increased with resistance, meaning that higher-resistance memristors caused greater output shifts for the same amount of injected current. From this, we derived a calibration equation to estimate the resistance from the slope: $R_{est.} = 1.501 \times |slope| - 1.47$.

We validated this model using test cases not included during training. For example, a 17 kΩ memristor was estimated as 17.4 kΩ when tested with 15 µA and 20 µA injection (2.35% error), but adding more test points (15–40 µA) improved the result to 16.94 kΩ with only 0.35% error. Similarly, for a 10 kΩ memristor, using high out-of-range test currents (50–100 µA) produced a 5.75% error, whereas staying within the trained range (12–40 µA) reduced the error to just 0.3%. These results show that using more test points and choosing fault injection values within the trained range significantly improves weight estimation accuracy. In the second phase, we explored whether injected currents could actively change the stored weights. Memristors exhibit a pinched hysteresis loop in their current–voltage characteristics, meaning the current response traces a looped path as voltage is swept forward and backward, pinching at the origin. This memory behavior is presented in Fig. 2. Simulations using the TEAM model showed that relatively high injected currents (100 µA to 1.2 mA) were sufficient to shift the internal state and alter the device resistance R(x). As shown in Fig. 3a, small faults like 10 µA have minimal effect, but higher magnitudes push the device toward the OFF state, increasing resistance. At an injection current of 1.2 mA, the memristor's resistance rises from 138 Ω to 336 Ω, an increase of ~140% over the baseline. The loop widens and the final state stabilizes at a new value, permanently modifying the stored weight.

TABLE I. CHANGE IN COLUMN OUTPUT CURRENT DUE TO INJECTED FAULT CURRENTS ACROSS DIFFERENT MEMRISTOR RESISTANCES

| Inj. Current (µA) | ΔI (µA) at 5 kΩ | ΔI (µA) at 10 kΩ | ΔI (µA) at 12 kΩ | ΔI (µA) at 15 kΩ | ΔI (µA) at 20 kΩ |
|---|---|---|---|---|---|
| 10 | 2.29 | 1.28 | 1.09 | 0.89 | 0.68 |
| 15 | 3.43 | 1.93 | 1.64 | 1.34 | 1.03 |
| 20 | 4.59 | 2.58 | 2.19 | 1.79 | 1.37 |
| 30 | 6.91 | 3.88 | 3.31 | 2.70 | 2.07 |
| 40 | 9.25 | 5.21 | 4.43 | 3.62 | 2.78 |

In summary, low-level laser-induced µA currents can expose weights for model extraction, while stronger injections can reprogram memristors by shifting their conductance.

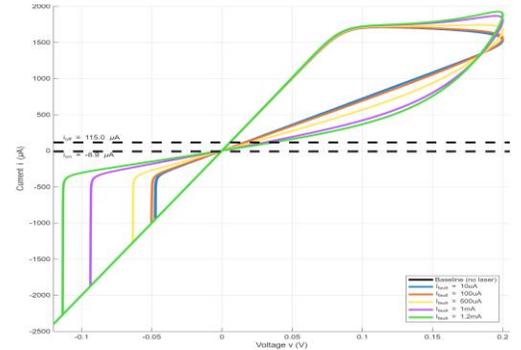

Fig. 2. Memristor I-V: Fault Current Sweep

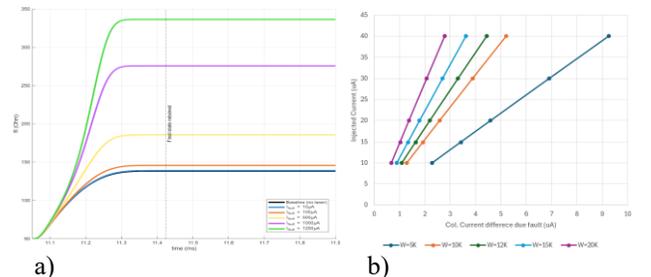

a)  b)

Fig. 3. a) Resistance R(x) under laser fault, b) Relation between weights when different fault current is injected

## References

[1] L. O. Chua, "Memristor—The Missing Circuit Element," *IEEE Transactions on Circuit Theory*, vol. 18, no. 5, pp. 507–519, 1971.

[2] D. B. Strukov, G. S. Snider, D. R. Stewart, and R. S. Williams, "The missing memristor found," *Nature*, vol. 453, pp. 80–83, 2008.

[3] M. F. Rahman and W. Burleson, "Integrated Security Mechanisms for Weight Protection in Memristive Crossbar Arrays," *2025 IEEE 36th International Conference on Application-specific Systems, Architectures and Processors (ASAP)*, Vancouver, BC, Canada, 2025, pp. 184-185, doi: 10.1109/ASAP65064.2025.00043.

[4] S. Kvatinsky, E. G. Friedman, A. Kolodny and U. C. Weiser, "TEAM: ThrEshold Adaptive Memristor Model," in *IEEE Transactions on Circuits and Systems I: Regular Papers*, vol. 60, no. 1, pp. 211-221, Jan. 2013, doi: 10.1109/TCSI.2012.2215714.

[5] K. Yamashita, B. Cyr, K. Fu, W. Burleson, and T. Sugawara, "Redshift: Manipulating Signal Propagation Delay via Continuous-Wave Lasers", *TCHES*, vol. 2022, no. 4, pp. 463–489, Aug. 2022, doi: 10.46586/tches.v2022.i4.463-489.